\documentclass[
    a4paper,
    10pt,
    superscriptaddress,
    amsmath, amssymb,
    aps,
    prx,
    reprint
]{revtex4-2}

\usepackage[top=1.5cm, bottom=1.5cm, left=1.5cm, right=1.5cm, bindingoffset = 0pt, footskip = 0.75cm, marginparwidth = 0pt, marginparsep=0pt]{geometry}

\UseRawInputEncoding
\usepackage{physics}
\usepackage{graphicx}
\usepackage{float}
\usepackage{dcolumn}
\usepackage[hidelinks]{hyperref}
\usepackage[capitalise]{cleveref}
\usepackage[normalem]{ulem}
\usepackage{tabularx}
\usepackage{longtable}
\usepackage{xcolor}
\usepackage{booktabs}
\usepackage{titlesec}

\makeatletter
\def\@fnsymbol#1{\ensuremath{\ifcase#1\or *\or *\dagger\or \ddagger\or
   \mathsection\or \mathparagraph\or \|\or **\or \dagger\dagger
   \or \ddagger\ddagger \else\@ctrerr\fi}}
\makeatother

\begin{document}

\title{High-Efficiency, Low-Loss Floquet-Mode Traveling-Wave Parametric Amplifier}

\newcommand{\RLE}{\affiliation{Research Laboratory of Electronics, Massachusetts Institute of Technology, \\ Cambridge, Massachusetts 02139, USA}}
\newcommand{\EECS}{\affiliation{Department of Electrical Engineering and Computer Science, Massachusetts Institute of Technology, \\ Cambridge, Massachusetts 02139, USA}}
\newcommand{\LL}{\affiliation{MIT Lincoln Laboratory, Lexington, Massachusetts 02421, USA}}
\newcommand{\PHYS}{\affiliation{Department of Physics, Massachusetts Institute of Technology, Cambridge, MA 02139, USA}}

\author{Jennifer Wang}
\thanks{These two authors contributed equally} \EECS\RLE

\author{Kaidong Peng}
\thanks{These two authors contributed equally} \email{kdpeng@mit.edu}  \EECS\RLE

\author{Jeffrey M. Knecht} \LL

\author{Gregory D. Cunningham} \EECS\RLE
\author{Andres E. Lombo} \EECS\RLE
\author{Alec Yen} \EECS\RLE
\author{Daniela A. Zaidenberg} \EECS\RLE

\author{Michael Gingras} \LL
\author{Bethany M. Niedzielski} \LL
\author{Hannah Stickler} \LL
\author{Katrina Sliwa} \LL
\author{Kyle Serniak} \RLE \LL
\author{Mollie E. Schwartz} \LL

\author{William D. Oliver} \EECS\RLE\PHYS

\author{Kevin P. O'Brien}
\email{kpobrien@mit.edu} \EECS\RLE

\date{\today}

\begin{abstract}
Advancing fault-tolerant quantum computing and fundamental science necessitates quantum-limited amplifiers with near-ideal quantum efficiency and multiplexing capability. However, existing solutions typically achieve one at the expense of the other. In this work, we experimentally demonstrate the first Floquet-mode traveling-wave parametric amplifier (Floquet TWPA), which achieves nearly quantum-limited noise performance, minimal dissipation, and broadband operation, breaking the presumption that broadband amplifiers introduce higher noise. We achieve a system measurement efficiency of $65.1\pm5.8\%$ when measuring a superconducting qubit, which to our knowledge is the highest-reported in a superconducting qubit readout experiment utilizing phase-preserving amplifiers. Our device exhibits $>20$-dB amplification over a $3$-GHz instantaneous bandwidth, $<\!0.5\,$-dB average in-band insertion loss, and the highest reported intrinsic quantum efficiency for a TWPA of $92.1\pm7.6\%$, relative to an ideal phase-preserving amplifier. Fabricated in a superconducting qubit process, these general-purpose Floquet TWPAs are suitable for fast, high-fidelity multiplexed readout in large-scale quantum systems and future monolithic integration with quantum processors. 

\end{abstract}

\maketitle

\section{Introduction}

Quantum-limited amplifiers enable faithful detection of single-photon-level signals by noisy real-world electronics \cite{caves_quantum_1982}. They are critical front-end hardware for various information-critical applications in quantum information science, metrology, and astronomy. Since their development, quantum-limited amplifiers have enabled fast and high-fidelity single-shot readout of quantum bit (qubit) states \cite{mallet_single_2009,lin_single_2013} and will continue to play an increasingly important role in the era of fault-tolerant quantum computing. Due to algorithmic complexity, finite qubit coherence, and imperfect hardware control \cite{krinner_realizing_2022}, large-scale quantum computers with potentially millions of physical qubits and active quantum error correction are needed to ensure fault-tolerant operations and empower real-world applications.

Surface codes \cite{kitaev_quantum_1997,bravyi_quantum_1998,dennis_topological_2002,kitaev_fault_2003} are leading quantum error correction (QEC) schemes due to their high error threshold, compatibility with 2D hardware architectures, and the availability of efficient error decoding algorithms \cite{edmonds_paths_1965,edmonds_maximum_1965}. In each QEC cycle, physical qubit errors are detected and decoded by sets of simultaneous stabilizer measurements on auxiliary qubits to extract the error syndromes of data qubits, without collapsing the underlying logical qubit state. For QEC to break even and improve the overall error rate, surface codes thus require quantum-limited amplifiers, which minimize both the readout errors and the error-correction cycle duration to reduce qubit decoherence. Additionally, the logical state preparation error also depends on the readout speed and fidelity of the corresponding initialization stabilizer measurements. Fast, high-fidelity readout is thus critical to the success of QEC, which imposes stringent performance and scalability requirements on the measurement hardware, especially the quantum-limited preamplifiers.

%---------- Fig 1 description ----------%
\begin{figure*}[ht]
\centering
\includegraphics[width=\linewidth]{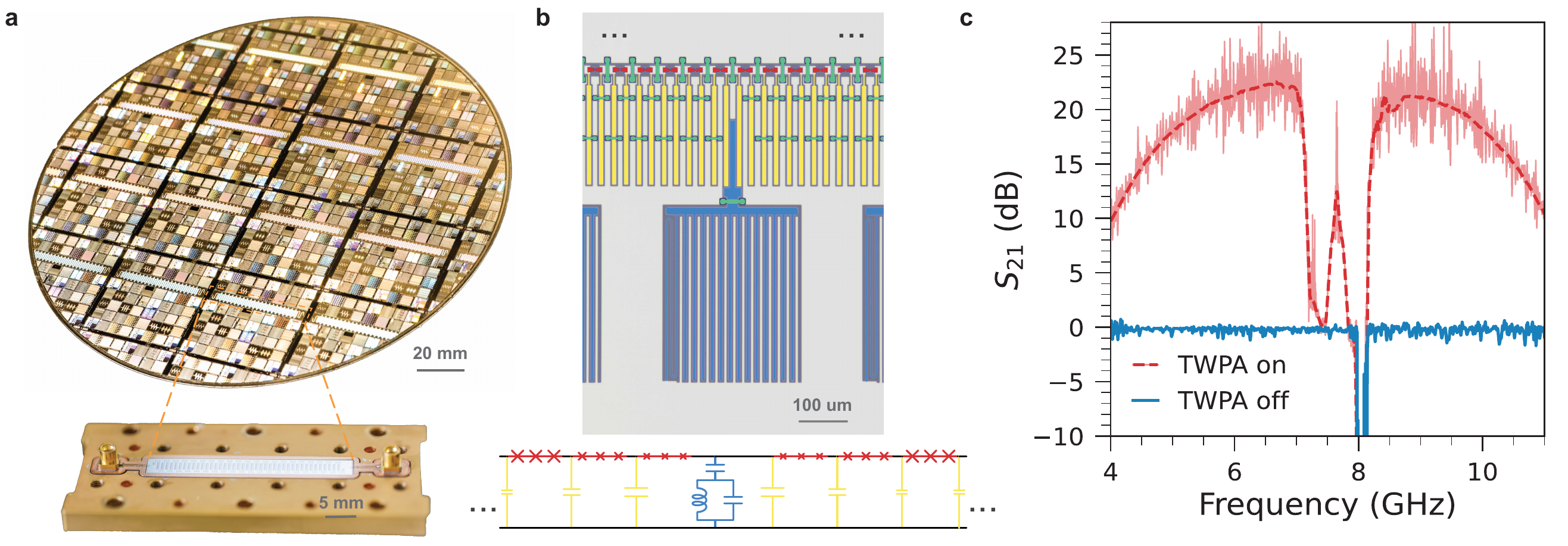}
\caption{ (a) Floquet TWPAs are fabricated on a 200-mm silicon wafer using a standard superconducting qubit process. The Floquet TWPA chips are $5 \times 40$ mm in size and are housed in a gold-plated package with SMPM connectors. (b) False-color micrograph and circuit schematic of the Floquet TWPA. Each LC ladder of the nonlinear transmission line consists of three Dolan-style Josephson junctions chained in series (red) and a coplanar stub capacitor (yellow) to ground (light gray). The junction sizes and coplanar stub lengths are varied as a function of position across the device. The capacitively coupled phase matching resonators (blue) are inserted every eight unit cells to phase match the parametric amplification process. The airbridges (green) are placed in every cell to connect the grounds on both sides of the junction chains. (c) Measured insertion loss (blue) and broadband parametric gain (light red) of the Floquet TWPA, both normalized to a through line with matching cable lengths in a microwave switch bank; the measurement setup is depicted in \cref{fig2}(a). The red dashed line is the parametric gain data smoothed with a Savitzky-Golay filter. }
\label{fig1}
\end{figure*}

Josephson traveling-wave parametric amplifiers (JTWPAs) \cite{macklin_a_2015,white_traveling_2015,planat_photonic-crystal_2020,ranadive_kerr_2022,gaydamachenko_an_2025} are promising candidates as quantum preamplifiers for large-scale fault-tolerant quantum computers. JTWPAs achieve near-quantum-limited noise performance over several gigahertz of instantaneous bandwidth and a high dynamic range ($-80$ to $-110\,$ dBm), capable of simultaneously reading out tens to hundreds of frequency-multiplexed qubits. However, JTWPAs introduce more noise than resonator-based amplifiers \cite{yurke_observation_1989,yamamoto_flux_2008,castellanos-beltran_widely_2007,mutus_strong_2014,roy_broadband_2015,white_readout_2023} due to non-negligible material loss and coherent sideband leakage \cite{macklin_a_2015,peng_floquet_2022}, leading to either degraded readout fidelity or extended measurement time.

Floquet-mode traveling-wave parametric amplifiers (Floquet TWPAs) \cite{peng_floquet_2022} have recently been proposed to overcome the quantum efficiency-bandwidth trade-off in conventional TWPAs. In comparison, Floquet TWPAs encode information in instantaneous Floquet modes rather than single-frequency modes to eliminate sideband leakage from parasitic nonlinear processes. Furthermore, Floquet TWPAs adiabatically mode-match the information-carrying Floquet modes to single-frequency modes for convenient usage, improved directionality, and insensitivity to out-of-band impedance mismatch, paving the way for direct integration with quantum processors by reducing the need for extraneous hardware such as microwave isolators. 

Experimentally realizing Floquet TWPAs with these advantages is challenging, as they require greater electrical length and are consequently more vulnerable to material dissipation than conventional TWPAs. TWPAs are often implemented with parallel-plate capacitors using lossy dielectrics such as silicon oxide \cite{macklin_a_2015,qiu_broadband_2022} and alumina \cite{planat_photonic-crystal_2020,ranadive_kerr_2022} ($\tan\delta \approx10^{-3}$). At this loss level, the intrinsic quantum efficiency of Floquet TWPAs would be limited by material loss and is predicted to decrease to $\sim90\%$ \cite{peng_x_2022} from $99.9\%$ when lossless, diminishing the performance advantages over the conventional amplifier paradigm. Since the completion of this manuscript, we have become aware of a recent preprint \cite{chang_josephson_2025}, which uses a superconducting-qubit fabrication process to implement a periodically-loaded JTWPA without the Floquet scheme.

In this work, we present the first experimental demonstration of a Floquet TWPA, fabricated in a superconducting qubit process. In addition to simpler fabrication and easier integrability with superconducting qubits, our implementation dramatically reduces material loss by more than $50\times$ relative to silicon oxide or alumina, using a single-layer defined distributed coplanar capacitors on a high-resistivity crystalline silicon substrate \cite{peng_quest_2023,chang_josephson_2025}. Overall, our implementation results in an effective loss tangent  $\tan\delta_{\mathrm{eff}}\approx 6\times10^{-5}$ and $<0.5\,$dB of average insertion loss within the entire 4-12 GHz frequency range, dominated by packaging and impedance mismatch and can be further minimized with qubit-amplifier integration in the future. Our Floquet TWPA achieves a peak gain exceeding $20\,$dB, several gigahertz of instantaneous $3$ dB bandwidth, and a $1$ dB compression point of $-106.5$ dBm at $21\,$dB gain, making it suitable for use in large-scale fault-tolerant quantum computers and various other information-critical applications. 

\section{Device Design}

Our Floquet TWPA is fabricated on a 200-mm high-resistivity ($>\!\!10\,\mathrm{k}\Omega\cdot$cm) float-zone silicon wafer using a standard superconducting qubit process \cite{Rosenberg_3d_2017,das_cryogenic_2018, Niedzielski_silicon_2019, rosenberg_3d_2019,Yost_solid_2020,mallek_fabrication_2021}, consisting of an aluminum ground plane, Josephson junctions \cite{dolan_offset_1977}, and aluminum air bridges \cite{chen_fabrication_2014} as shown in Fig.~\ref{fig1}(a). Meandered with a $5\times40$ mm chip footprint, the Floquet TWPA consists of 3008 sets of three-Josephson-junction chains interleaved by open-ended coplanar waveguide stubs, which function as low-loss distributed capacitors to ground (see \cref{app:stubs} for further details). Capacitively coupled lumped-LC resonators centered at $\sim8\,$ GHz are inserted every 8 unit cells to phase-match the desired four-wave-mixing parametric amplification process \cite{obrien_resonant_2014}. The lengths of the Josephson junctions and the coplanar stubs in each cell are jointly varied across the device to control the instantaneous nonlinearity and implement the adiabatic Floquet scheme \cite{peng_floquet_2022} while maintaining $\sim50\,\Omega$ impedance matching between cells. The resulting junction critical currents range from $4.62\,\mu$A in the middle to $13.1\,\mu$A at the ends, chosen to balance Floquet mode matching at the interfaces and $\sim-100\,$dBm P$_{1\mathrm{dB}}$ gain compression for multiplexing tens of qubits. 

As shown in Fig.~\ref{fig1}(c), The total device insertion loss averages $<0.5\,$dB within the full 4-12 GHz band. Because dielectric loss $A_{\mathrm{dielectric}}[dB] \propto k(\omega)\tan\delta \propto \omega$ is linearly proportional to wavevector $k(\omega$) and signal frequency $\omega$ on the dB scale and our device exhibits a near-zero slope, we attribute the remaining $\sim0.11\,$dB frequency-independent insertion loss (see \cref{app:devicefit} for further details) to microwave packaging and impedance mismatch, which can be optimized. In addition to minimal insertion loss, our Floquet TWPA exhibits $>20\,$dB parametric gain over an instantaneous bandwidth of $3\,$GHz as shown in Fig.~\ref{fig1}(c). We highlight that the parametric gain reported here in Fig.~\ref{fig1}(c) is relative to a through line and not to the undriven case. The device is measured in a realistic measurement environment with directional couplers, circulators, and no attenuators which would otherwise improve impedance matching conditions and gain performance for TWPAs.

\section{Intrinsic Amplifier Quantum Efficiency}

%---------- Fig 2 description ----------%
\begin{figure}[]
\centering
\includegraphics[width=\columnwidth]{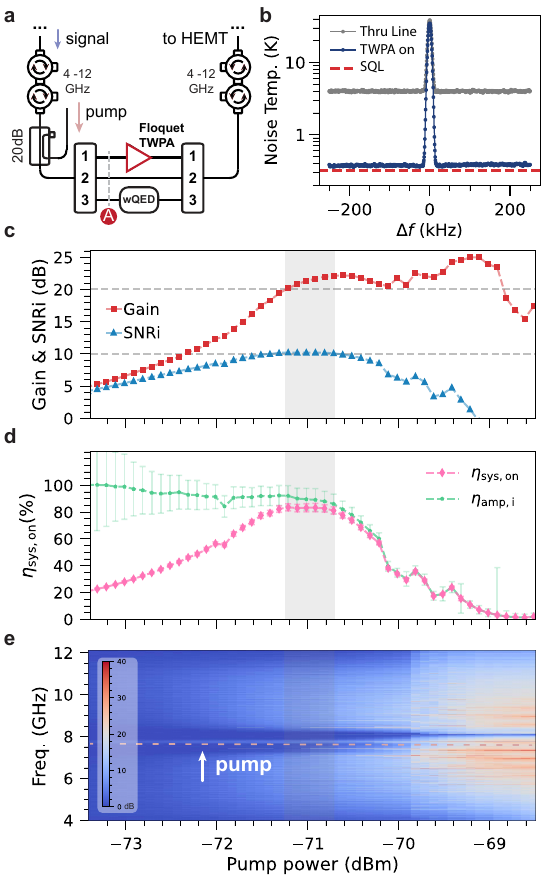}
\caption{(a) Measurement setup for performing wQED power calibration and characterizing the intrinisic amplifier quantum efficiency. The plane of absolute power reference is placed right before the wQED and the Floquet TWPA within the switch bank, labeled by the red letter \textbf{A}. (b) Noise temperatures measured at the optimal pump configuration with the Floquet TWPA turned on (blue) and with it switched out (gray), respectively. The red dashed line corresponds to the standard quantum limit (SQL). Data is taken at a signal frequency of 6.59 GHz. (c) The measured gain and SNRi (see main text for definition) of the Floquet JTWPA as a function of the applied pump power at pump frequency 7.71 GHz. The optimal pump operating region is highlighted in light gray, within which the maximal SNRi of 10.23 dB occurs at a parametric gain of 20.18 dB. (d) The extracted system measurement efficiency $\eta_{\mathrm{sys,on}}$ (pink) and intrinsic amplifier quantum efficiency $\eta_{\mathrm{amp,i}}$ (green), respectively, with maximal values of  $83.6 \pm 2.6\%$ and $92.1 \pm 7.6\%$. (e) The measured noise power spectra, normalized by the small-pump-power noise powers, as a function of the applied pump power on the decibel scale. The dashed line at 7.71 GHz corresponds to the applied pump as indicated by the white arrow and text. An abrupt noise floor rise is visible at around $-69.85\,$dBm, indicative of the onset of self-oscillations.}
\label{fig2}
\end{figure}

We evaluate the intrinsic noise performance of our Floquet TWPA using a waveguide quantum electrodynamics (wQED) method \cite{mirhosseini_cqed_2019,kannan_generating_2020,qiu_broadband_2022,kannan_wqed_2023}. The wQED device, consisting of a qubit symmetrically coupled to a transmission line, provides an absolute power reference at the cryogenic stage at the qubit frequency for noise characterization. At weak coherent drive powers ($\vert \alpha \vert^2 \ll 1$ where $\alpha$ is the coherent state amplitude), the qubit absorbs a single photon from the drive field and emits a photon bidirectionally with a $\pi$ phase shift, thereby perfectly reflecting the single photon through constructive interference in the backward direction and destructive interference in the forward direction \cite{qiu_broadband_2022,kannan_wqed_2023}. At higher drive powers, the transmission increases and approaches unity. The power-dependent transmission can thus be fitted to extract the desired absolute power reference. 

\Cref{fig2}(a) shows the experimental setup used to characterize the intrinsic amplifier quantum efficiency. The Floquet TWPA and the wQED device are placed in parallel inside a pair of multi-throw microwave switches, and the pump is combined with the input signal tone through a $20\,$dB directional coupler before the switch bank. The wQED device thus provides a power reference at the reference plane \textbf{A}, and we use nominally identical microwave cables with matching lengths and similar PCB packaging to minimize line differences between the different switch bank paths. After the switch bank, the signal is routed to a double-junction circulator, a superconducting NbTi cable, and a high-electron-mobility transistor amplifier (HEMT) for further amplification at 4K. We choose to operate and obtain an absolute power at the flux sweet spot qubit frequency of 6.59 GHz, which is also near the center of the TWPA high gain band.

In \cref{fig2}(c), we plot the measured parametric gain (red) and the associated SNR improvement (blue),  $\mathrm{SNRi} \equiv \mathrm{SNR}_{\mathrm{twpa,on}} / \mathrm{SNR}_{\mathrm{thru}}$, as a function of the input pump power at the pump frequency $7.71\,$GHz. Instead of the common definition of SNRi relative to an undriven device, we define the SNRi relative to a through line, because this fairly represents the SNR improvement from adding the amplifier into the measurement chain for devices with insertion loss. The parametric gain increases monotonically with increasing pump power and reaches a local peak of $22.17\,$dB at a pump power of $\sim-70.5\,$dBm before decreasing. The SNRi follows a similar trend and peaks at $10.23\,$dB with a parametric gain of $20.19\,$dB, occurring at a gain just $1.98\,$dB before the peak gain at the pump power $\sim-71.2\,$dBm. The close proximity of the gain and the SNRi optima in our Floquet TWPA shows an improvement over conventional TWPAs, as predicted in \cite{peng_floquet_2022} from more efficient sideband suppression. 

In \cref{fig2}(d), we show the corresponding system measurement efficiency %\eta_{sys} $ 
\begin{equation}
\eta_{\mathrm{sys, on}} = \frac{\hbar \omega}{k_B T_{\mathrm{sys, on}} }\label{eq:qesysdef}
\end{equation} 
and the deduced phase-preserving intrinsic amplifier quantum efficiency (see \cref{app:intinsicqe} for further details)
\begin{equation}
\eta_{\mathrm{amp,i}}(G)\! = \! \!{\left(\frac{2}{\eta_{\mathrm{sys,on}}} - \frac{2}{G\eta_{\mathrm{sys,off}}} + \frac{1}{G}\right)}^{-1}\!\!\Big/ \, \eta_{\mathrm{ideal}}(G),
\label{eq:qeintrinsicdef}
\end{equation}
as calculated by applying the aforementioned wQED power calibration on the noise measurement. In \cref{eq:qeintrinsicdef}, G is the amplifier power gain (in linear units), $\eta_{\mathrm{ideal}}(G)=1 /(2-1/G)$ \cite{peng_floquet_2022} is the quantum efficiency of an ideal phase-preserving quantum-limited amplifier at gain G, $\eta_{\mathrm{sys,off}}$ refers to the system measurement efficiency when the Floquet TWPA is turned off. We note that in this definition, the intrinsic amplifier quantum efficiency is normalized by that of an ideal phase-preserving quantum-limited amplifier, with $100\%$ corresponding to the standard quantum limit. 

Between pump powers of $-71.25\,$ dBm and $-70.7\,$ dBm exists a region (highlighted in light gray) in which the system measurement efficiency exceeds $80\%$ and the parametric gain exceeds $20\,$dB concurrently. The optimal system measurement efficiency reaches $\eta_{\mathrm{sys, on}} = 83.6 \pm 2.6\%$, for which the downstream amplification chain (including the HEMT) contributes to $9.22\pm1.05\%$ of the inefficiencies. This is among the highest reported system measurement efficiencies \cite{chang_josephson_2025}, and we note that in contrast to Ref.$\,$\cite{chang_josephson_2025}, our system measurement efficiency is measured without any attenuators immediately before or after the amplifiers, which would otherwise improve impedance matching conditions. At low parametric gain, the intrinsic amplifier quantum efficiency approaches the quantum limit, within measurement uncertainty. At the same operating point for optimal system measurement efficiency and $>20\,$dB gain, the intrinsic amplifier quantum efficiency is $\eta_{\mathrm{amp,i}} = 92.1 \pm 7.6\%$, to our knowledge the highest reported value in TWPAs \cite{macklin_a_2015,white_traveling_2015,planat_photonic-crystal_2020,ranadive_kerr_2022,ranadive_traveling_2024}. \Cref{fig2}(b) shows the system noise temperature of $378\pm25\,$mK and $3.99\pm0.25\,$K when the Floquet TWPA is turned on (blue) and switched out (gray), respectively, at the optimal pump operating point discussed above.

For the discussions above, the system noise is referred to the input of the wQED device/the Floquet TWPA as shown in \cref{fig2}(a). The choice of this plane of reference \textbf{A} is suitable for extracting the intrinsic amplifier quantum efficiency, as is the central focus of this section. Nonetheless, we note that the attained high system measurement efficiency $\eta_{\mathrm{sys,on}}= 83.6 \pm 2.6\%$ shows great promise for minimizing the insertion loss between the preamplifier and the quantum processor. In addition, the above analysis sheds light on the limiting factors at high system measurement efficiencies. As preamplifiers continue to advance in performance and become tightly integrated with qubits, noise from downstream measurement chain will soon dominate and limit the attainable system measurement efficiency to be under $\sim\!90\%$, even at a $20\,$dB preamplifier gain. Near-ideal quantum measurements require preamplifiers with higher gain, HEMTs with improved noise performance, or a second stage of near-quantum-limited amplifiers following preamplification. 

Above a pump power of $-70\,$dBm, there exists a region where the parametric gain increases to $\approx25\,$dB while the SNRi decreases significantly. The SNRi becomes negative around $-69\,$dBm and parametric gain decreases again. \Cref{fig2}(e) shows the corresponding noise power rise spectra, normalized by the small-pump-power noise powers, as a function of frequency and pump power. We observe that these high gain regions with poor noise performance are correlated with uniformly spaced noise peaks, which form symmetrically about the dispersion feature and become increasingly more prominent at higher parametric gain. At a pump power of $\sim-69.85\,$dBm, beyond the optimal operating point, an abrupt noise floor rise occurs across the entire spectrum, indicative of the onset of parametric self-oscillations, and consistent with the sudden increase in measured parametric gain. We hypothesize that the noise peaks are caused by self-oscillations originating in an effective cavity in the middle of the Floquet TWPA, due to the varying phase-matching resonator frequencies. The self-oscillation effect is also exacerbated by the minimal insertion loss and imperfect critical current density targeting from fabrication ($\sim\!\!17.7\%$ difference between measured and designed $J_c$).

\section{System Measurement Efficiency for Qubit Readout \label{sec:sysmeaseff}}

%---------- Fig 3 description ----------%
\begin{figure}
\centering
\includegraphics[width=\columnwidth]{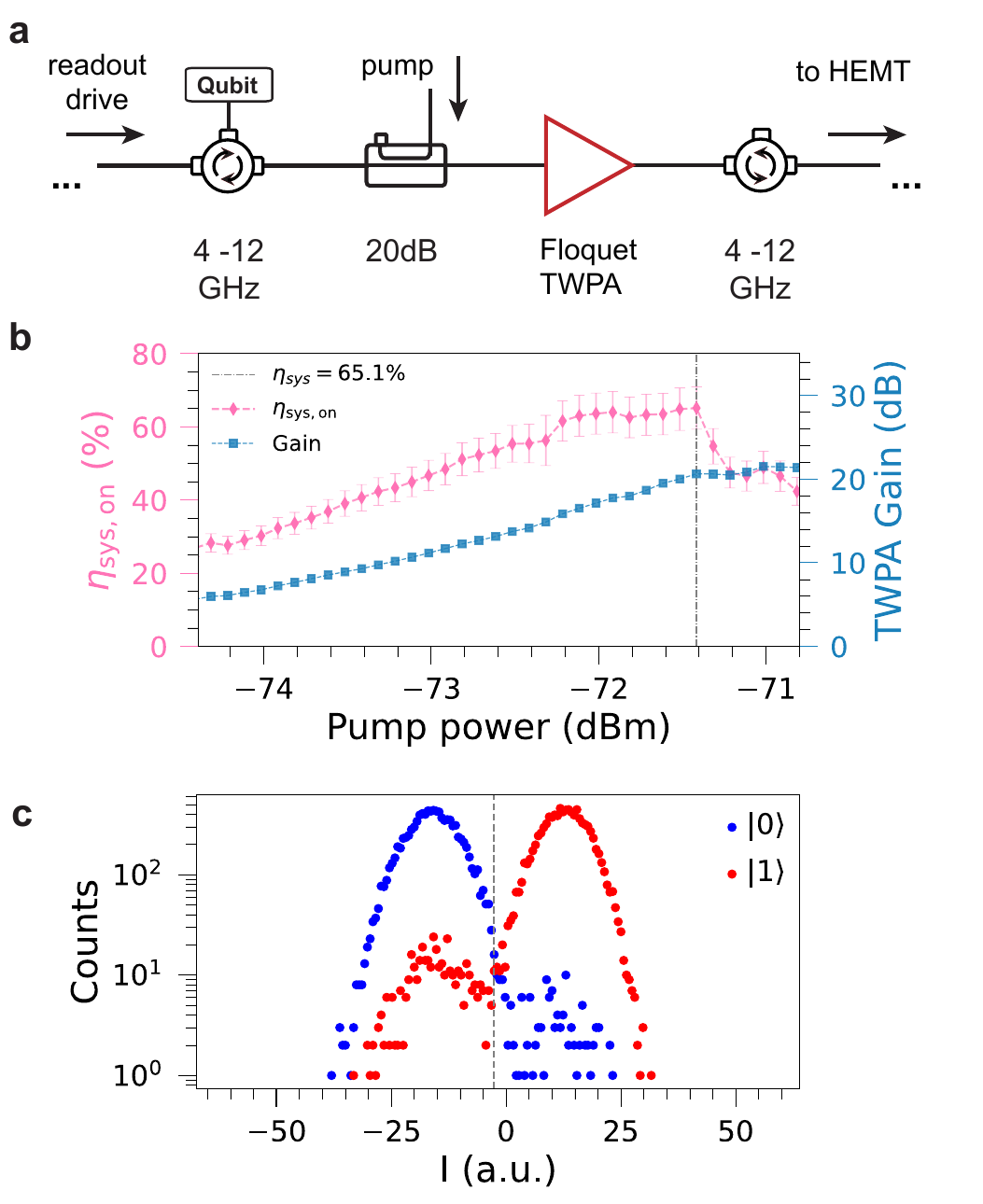}
\caption{(a) Measurement setup for qubit readout, featuring a transmon qubit in reflection followed by the Floquet TWPA. (b) Using measurement-induced dephasing of the qubit, system efficiency is calculated to be $65.1\% \pm 5.8\%$ at max (pink dots). At the same pump power, the parametric gain is $20.6$ dB (blue squares). (c) Qubit readout with the JTWPA pump on, resulting in an average readout fidelity of $97.23\%$ over an integration time of 0.5 $\mu s$, with the threshold between the ground and excited states defined by the gray dashed line.} 
\label{fig3}
\end{figure}

We now use the Floquet TWPA in a qubit readout experiment. As shown in \cref{fig3}(a), we perform a reflection measurement on a transmon qubit separated from the TWPA by a single-junction circulator and a $20$-dB directional coupler. A circulator after the Floquet TWPA mitigates reflections from subsequent components and reduces thermal noise from higher temperature stages.

To characterize the system measurement efficiency for qubit readout, we perform an in-situ power calibration using the measurement-induced dephasing method \cite{gambetta_mid_2006,macklin_a_2015} on the transmon and readout resonator. A coherent tone populates the resonator, causing an AC Stark shift and measurement-induced dephasing of the qubit. We obtain an absolute power reference at the input of the qubit readout resonator by fitting to the functional dependence of these rates on the resonator drive power and detuning (see \cref{app:qubitcal} for further details). 

We perform noise spectral power density measurements and extract the maximum system efficiency to be $65.1\% \pm 5.8\%$, at $20.6$ dB gain, depicted in \cref{fig3}(b). This measurement efficiency is significantly higher than previously reported qubit-readout-chain measurement efficiencies using TWPAs of approximately $20\%$ to $50\%$ \cite{macklin_a_2015, bultink_2018, heinsoo_2018, peronnin_2020, andersen_2019, andersen_2020, planat_photonic-crystal_2020, ranzani_kinetic_2018}, and is among the best achieved by phase-preserving Josephson parametric amplifiers (JPAs), ranging from $55\%$ to $63\%$ for similar readout chains \cite{kaufman_simple_2024, kaufman_josephson_2023, bothara_tantalum_2025, kutlu_fluxjpa_2021, uchaikin_capp_2024, abdo_2019, abdo_2021}. We note that when operated in phase-sensitive configurations, parametric amplifiers can provide higher system efficiencies \cite{eddins_2018, lecocq_2021, rosenthal_efficient_2021, walter_rapid_2017}. With the TWPA pump on, we obtain representative single shot readout histograms in Fig.~\ref{fig3}(c) and measure a $T_1$-limited readout fidelity of $97.23\%$ on average, with a readout length of $0.5\,\mathrm{\mu s}$. Overall, the Floquet TWPA provides an attractive balance of wide bandwidth, high gain, and low noise for applications such as multiplexed readout of superconducting or spin qubits \cite{heinsoo_2018,andersen_2019,andersen_2020,krinner_realizing_2022,elhomsy_spin_2023}. 

\section{Dynamic Range}

%---------- Fig 4 description ----------%
\begin{figure}
\centering
\includegraphics[width=\columnwidth]{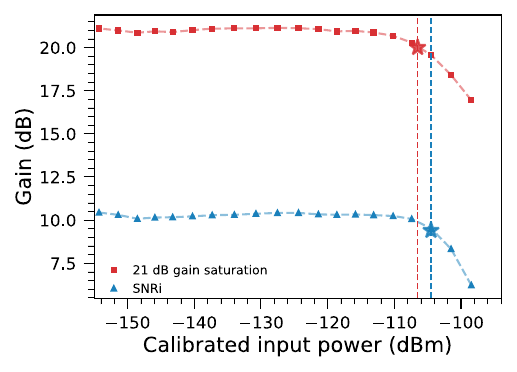}
\caption{Dynamic range of the Floquet JTWPA at a signal frequency of 6.59 GHz shows $P_{1\mathrm{dB}}$ at $-106.5\,$dBm (red star). The $1$-dB SNRi compression point is at $\sim-104\,$dBm (blue star), which is $2.5\,$dB higher than the gain compression point, indicating that the Floquet TWPA achieves improved noise performance compared to conventional counterparts.}
\label{fig4}
\end{figure}

In \cref{fig4}, we plot the measured parametric gain and SNRi at $6.59\,$GHz as a function of the calibrated input signal power. At $21$-dB gain, the $1$-dB compression point is $P_{1\mathrm{dB}} = -106.5\,$dBm, as indicated by the red star marker. Remarkably, if we similarly define a $1$-dB SNRi compression point as the signal power at which the SNRi lowers by $1\,$dB from the small-signal-power limit, the $1$-dB SNRi compression point occurs at a signal power $2.5\,$dB higher than the $P_{1\mathrm{dB}}$ gain compression point, as indicated by the blue star marker in \cref{fig4}. This is in stark contrast to conventional TWPAs, for which the SNRi compression occurs before the signal power reaches the gain compression point \cite{Remm_2023}. The improved noise performance at high signal power yields another advantage over conventional TWPAs, potentially attributed to its efficient sideband leakage suppression \cite{peng_x_2022}.

The measured $-106.5\,$dBm gain compression is on par with those of the experimentally realized conventional JTWPAs ($-110\,$dBm to $-95\,$dBm) reported in literature \cite{macklin_a_2015, planat_photonic-crystal_2020, qiu_broadband_2022, ranadive_kerr_2022,gal_gaincompression_2025}, showing promise for broadband photon emission detection and multiplexed qubit readout applications \cite{krinner_engineering_2019,Remm_2023,elhomsy_spin_2023}. Nonetheless, we note that there is still room for improvement of $P_{1\mathrm{dB}}$. The predicted gain compression point, calculated using the pump depletion model \cite{kylemark_2006} indicates a $P_{1\mathrm{dB}}$ value $\sim\!5.5\,$dB higher than the measured value. We hypothesize that the cause for the lower measured $P_{1\mathrm{dB}}$ is not signal sideband leakage as in conventional JTWPAs, evidenced by the later-occurring SNRi compression, but rather by the additional pump depletion from pump intermodulation byproducts and self-oscillation formation inside the Floquet TWPA, as discussed with \cref{fig2}(e). In future work, this parametric oscillation can be alleviated by tighter fabrication tolerance and better compensation of small resonance variations from varying capacitive couplings, which will increase parametric oscillation threshold and potentially lead to a $P_{1\mathrm{dB}}$ closer to theory. 
Furthermore, this particular Floquet TWPA device has a measured critical current density $J_c$  $17.7\%$ smaller than the design value. With better $J_c$ targeting, $P_{1\mathrm{dB}}$ will not only improve from higher junction critical current and input pump power, but also from reduced impedance mismatch and the resulting higher parametric oscillation threshold. In addition, $P_{1\mathrm{dB}}$ can also be  improved with longer junction chains with proportionally larger critical currents.

\section{Conclusion and Outlook}

In this work, we have presented the first experimental demonstration of a Floquet TWPA, featuring $>\!20\,$dB parametric gain over $3\,$GHz of instantaneous bandwidth. Implemented using a standard superconducting qubit fabrication process, this Floquet TWPA has an average insertion loss of $<\!0.5\,$dB within the $4$-$12\,$GHz band, which is among the lowest insertion losses reported for a JTWPA. At $>\!20\,$dB parametric gain, we measure a highest reported $92.1 \pm 7.6\%$ intrinsic amplifier quantum efficiency among phase-preserving TWPAs. When used in a superconducting qubit measurement setup as a general-purpose preamplifier, this Floquet TWPA enables a system measurement efficiency of $65.1 \pm 5.8\%$, the highest reported in a superconducting qubit readout experiment with a phase-preserving amplifier to the best of our knowledge.

Further improving system measurement efficiency towards ideal quantum measurements requires tighter integration of qubits and quantum-limited amplifiers to minimize insertion loss before preamplification. In future experiments, our qubit-process-compatible Floquet TWPA may be monolithically integrated with quantum processors to attain a projected $>80\%$ measurement efficiency, empowering fundamental physics research and fault-tolerant quantum computing.

\section*{Acknowledgments}
This work is supported in part by the Intelligence Advanced Research Projects Activity (IARPA), AWS Center for Quantum Computing, and Army Research Office under award No. W911NF-23-1-0045. We gratefully acknowledge MIT Lincoln Laboratory for fabricating the JTWPA through the support of Air Force Contract No. FA8702-15-D-0001.  Any opinions, findings, conclusions or recommendations expressed in this material are those of the author(s) and do not necessarily reflect the views of the Intelligence Advanced Research Projects Activity or the U.S. Air Force. J. Wang, A. Yen, and D. Zaidenberg acknowledge support from the NSF GRFP. We thank Aziza Almanakly and the MIT EQuS group for generously providing the wQED device used for power calibration. \\

K.P. and K.P.O. developed the theory and numerical simulation tools. K.P. designed the device and performed electromagnetic simulations with help from J.W.. K.P., K. Serniak, M.E.S., W.D.O. and K.P.O. conceived of fabricating TWPAs in a qubit process. J.K., M.G., B.M.N., H.S., K. Sliwa, K. Serniak, and M.E.S., fabricated the device. J.W. and K.P. performed the experiments and data analysis with help from G.D.C. and A. Y.. J.W., K.P., A.E.L., and D.A.Z. contributed to the experimental setup. K.P. and J.W. wrote the manuscript with feedback from all authors. K.P.O. supervised the project.

\appendix

\section{High-Q Distributed Coplanar Capacitors \label{app:stubs}}

%---------- Fig 5 description ----------%
\begin{figure}[H]
\centering
\includegraphics[width=0.8\columnwidth]{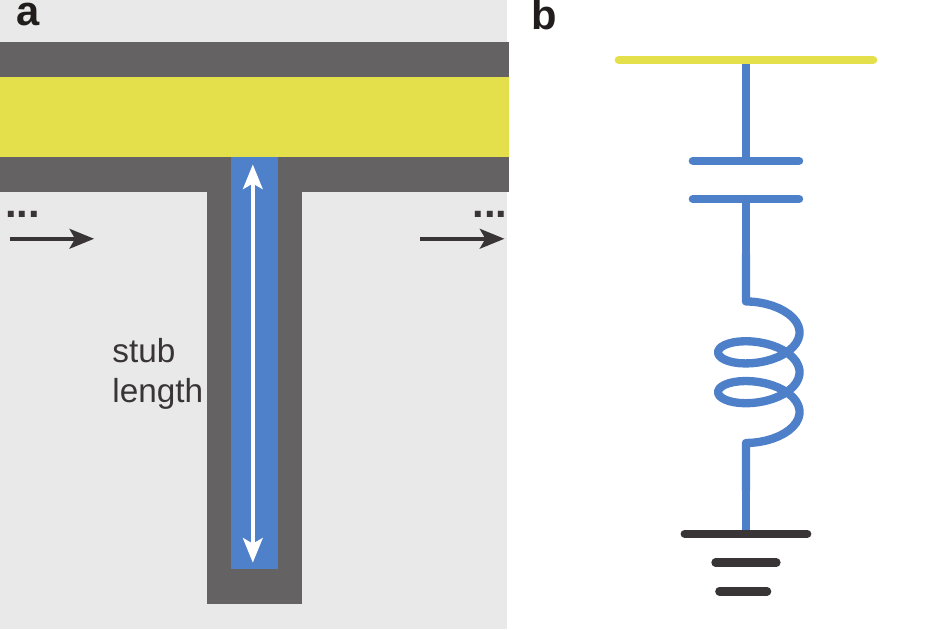}
\caption{The geometric illustration (a) and the equivalent circuit model (b) of the high-Q distributed coplanar capacitor implementation used in this work. The open-ended coplanar stub (blue) effectively forms a combination of a capacitance and inductance in series from the center trace (yellow) to the surrounding ground plane (light gray), and the b}
\label{fig:supp-fig5}
\end{figure}

\Cref{fig:supp-fig5} illustrates our low-loss capacitor implementation with open-ended coplanar waveguide stubs. From microwave theory, open-ended stubs have an input impedance 
\begin{equation}
    Z_{\mathrm{stub}}(\omega)=-j Z_{\mathrm{stub}} \cot \left(\frac{\omega}{2 \omega_{\mathrm{qw}}} \pi\right), \label{eq:zstub}
\end{equation}
in which $Z_{\mathrm{stub}}$ is the characteristic impedance of the coplanar stub, and $\omega_{\mathrm{qw}}=(\pi\nu)/(2l)$ is the quarter-wave resonant frequency of the coplanar stub. In the expression of  $\omega_{\mathrm{qw}}$, $\nu$ and $l$ are the wave velocity and the physical length of the coplanar stub, respectively. In the limit where the frequency of interest $\omega$ is much smaller than the resonant frequency $\omega_{\mathrm{qw}}$, the input impedance of an open-ended stub can be well approximated by a series combination of capacitor $C_{\mathrm{stub}}$ and inductor $L_{\mathrm{stub}}$ to ground:

\begin{equation}
\begin{aligned}
    Z_{\mathrm{stub}}(\omega) & \approx \frac{1}{j \omega\left(\pi / 2 \omega_{\mathrm{qw}}  Z_{\mathrm{stub}}\right)}+j \omega\left(\frac{\pi Z_{\mathrm{stub}}}{6 \omega_{\mathrm{qw}}}\right) \\
    &= \frac{1}{j\omega C_{\mathrm{stub}}} + j\omega L_{\mathrm{stub}}
\end{aligned}
\end{equation}
using series expansion, in which 
\begin{equation}
\begin{aligned}
    C_\mathrm{stub}&=\frac{\pi}{2\omega_{\mathrm{qw}}Z_{\mathrm{stub}}} = \frac{l}{\nu Z_{\mathrm{stub}}} \quad \mathrm{and} \\
    L_\mathrm{stub}&=\frac{\pi Z_{\mathrm{stub}}}{6\omega_{\mathrm{qw}}}=\frac{Z_{\mathrm{stub}}l}{3\nu}.
\end{aligned} \label{eq:LCstubexpr}
\end{equation}

 The inductance contribution $j\omega L_{\mathrm{stub}}$, a higher-order expansion term, is only significant at large $\omega/\omega_{qw}$ values. For the Floquet TWPA design considered in this work, the capacitance term alone is adequate. \Cref{eq:LCstubexpr} shows that the effective capacitance $C_\mathrm{stub}$ is linearly proportional to the stub length $l$, allowing the desired ground capacitance profile $C_g(x)$ to be implemented by controlling the lengths and the resonant frequencies of the stubs.

\section{Device Simulation and Measurement Analysis \label{app:devicefit}}

%---------- Fig 6 description ----------%
\begin{figure*}[ht]
\centering
\includegraphics[width=\linewidth]{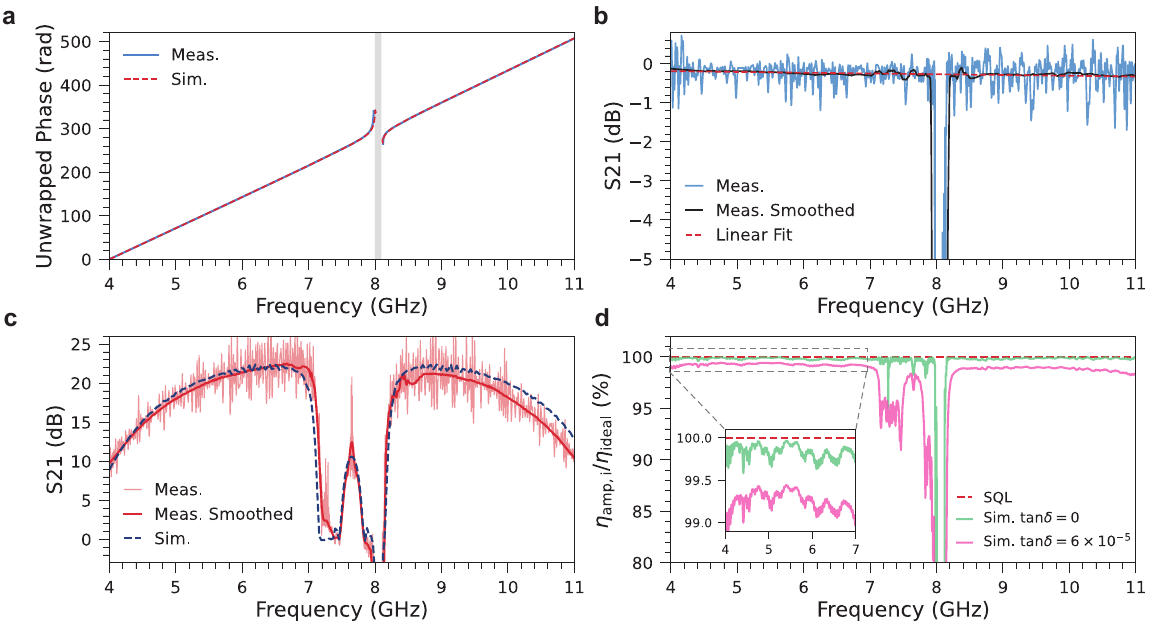}
\caption{ (a) Measured (blue, solid line) and simulated (red, dashed line) phase of the Floquet TWPA as a function of the signal frequency. The gray-shaded region at $\sim8\,$GHz corresponds to the phase-matching resonator bandgap, within which the transmission level is too low to be resolved. (b) Zoomed-in view of the measured (blue) insertion loss of the Floque TWPA relative to a through line with matching cable lengths. 
The Savitzky-Golay filter with a  501-point window is applied on the same measurement data and plotted in solid black line. The effective loss tangent from linear fitting the insertion loss (red, dashed line) is $\tan\delta_{\mathrm{eff}} = 6\times10^{-5}$. (c) Measured (red) and simulated (dashed, blue line) parametric gain of the Floquet TWPA. The Savitzky-Golay filter with a 251-point window is applied on the same measured gain data and plotted in solid red line for visual guidance. (d) The simulated amplifier intrinsic quantum efficiency $\eta_{\mathrm{amp,i}}/\eta_{\mathrm{ideal}}$ of our Floquet TWPA design assuming zero dielectric loss (green) or the fitted device loss tangent $\tan\delta_{\mathrm{eff}} = 6\times10^{-5}$ (pink) from (b). The standard quantum limit corresponds to $100\%$ and is plotted in red dashed line. The inset shows a zoomed-in view of the simulated intrinsic quantum efficiency values between 4 to 7 GHz.}
\label{fig:supp-fig6}
\end{figure*}

We perform nonlinear circuit simulation and measurement fitting on the Flqouet TWPA device of this work using JosephsonCircuits.jl \cite{peng_x_2022}, an open-source Julia simulation framework developed in our group that provides orders of magnitude speedup over time-domain-based simulation methods. In \cref{fig:supp-fig6}(a), we plot the measured and fitted unwrapped phase of the signal transmission relative to a through line with matching cable lengths in the same microwave switch banks. The gray-shaded region at $\sim8\,$GHz corresponds to the phase-matching resonator bandgap, which effectively functions as a narrow-band notch filter, blocking majority of the signal transmission and preventing phase to be accurately resolved inside the gap. To fit for the phase shift, we assume the deisgn capacitance values and the phase-matching-resonator frequency extracted from electromagnetic simulations, take into account the designed junction areas, and fit for the global critical current density as a free parameter. We achieve excellent agreement with the measured phase using a critical current density value of $J_c=10.7\,\mu\mathrm{A}\cdot\mu \mathrm{m}^{-2}$, consistent with the room-temperature DC probing value of $11.47\,\mu\mathrm{A}\cdot\mu \mathrm{m}^{-2}$. It is $17.7\%$ lower than the designed target of $13\,\mu\mathrm{A}\cdot\mu \mathrm{m}^{-2}$, which introduces unwanted impedance mismatch and lowers the parametric oscillation threshold discussed in \cref{fig2}(e).

In \cref{fig:supp-fig6}(b), we plot a zoomed-in view of the measured device insertion loss of \cref{fig1}(c) in the main text. We 
smooth the measurement data by applying a second-order Savitzky-Golay filter with a 501-point window, and perform a linear fit of the insertion loss in dB-scale versus signal frequency (red, dashed line). This is because the conductor loss of aluminum traces is minimal when superconducting, and the dielectric loss contribution ($\mathrm{Attn}_{\mathrm{dielectric}}$, in dB scale) dominates and scales linearly with input frequency, as illustrated in \cref{eq:tandfreqscaling}.

\begin{equation}
\begin{aligned}
\mathrm{Attn}_{\mathrm{dielectric}} & (\mathrm{dB}) = \mathrm{dB}\left(\left|e^{-\int_{x=0}^L \gamma(x) d x}\right|\right) 
 \\
 &=-\int_{x=0}^L k(x) \tan \delta  d x \propto\  \omega \times\tan \delta. 
\end{aligned} \label{eq:tandfreqscaling}
\end{equation}

From the linear fit, we extract a frequency-dependent insertion loss component of $0.11\,$dB and attribute it to impedance match and device packaging. The fitted slope gives an effective loss tangent of $\tan\delta_{\mathrm{eff}} \approx 6\times10^{-5}$, 50 times smaller than TWPAs implemented with silicon oxide or alumina parallel-plate capacitors.

In \cref{fig:supp-fig6}(c), we plot the measured (and smoothed) parametric gain curve of our device. Using the fitted circuit parameters and effective dielectric loss tangent from fitting \cref{fig:supp-fig6}(a) and (b), we fit for simulated parametric gain using the drive pump current as the only free parameter. Using a pump current of $I_p=0.392I_0$, where $I_0$ is the minimal junction critical current in the middle of the TWPA, we achieve excellent agreement with the measured gain spectrum. 

Finally, we plot the simulated intrinsic amplifier quantum efficiency of our design assuming a zero insertion loss (green) or the fitted effective dielectric loss tangent $\tan\delta_{\mathrm{eff}} = 6\times10^{-5}$. The dip seen in the simulated curve assuming realistic device loss just over $7\,$ GHz is due to the low parametric gain and finite insertion loss there. In both cases, the simulated amplifier efficiency values are typically higher than $99\%$, showcasing the advantages of the Floquet TWPA paradigm and the effectiveness of our low-loss, qubit-compatible fabrication process.

\section{Waveguide Quantum Electrodynamics (wQED) Power Calibration \label{app:wqed}}

%---------- Fig 7 description ----------%
\begin{figure}
\centering
\includegraphics[width=\columnwidth]{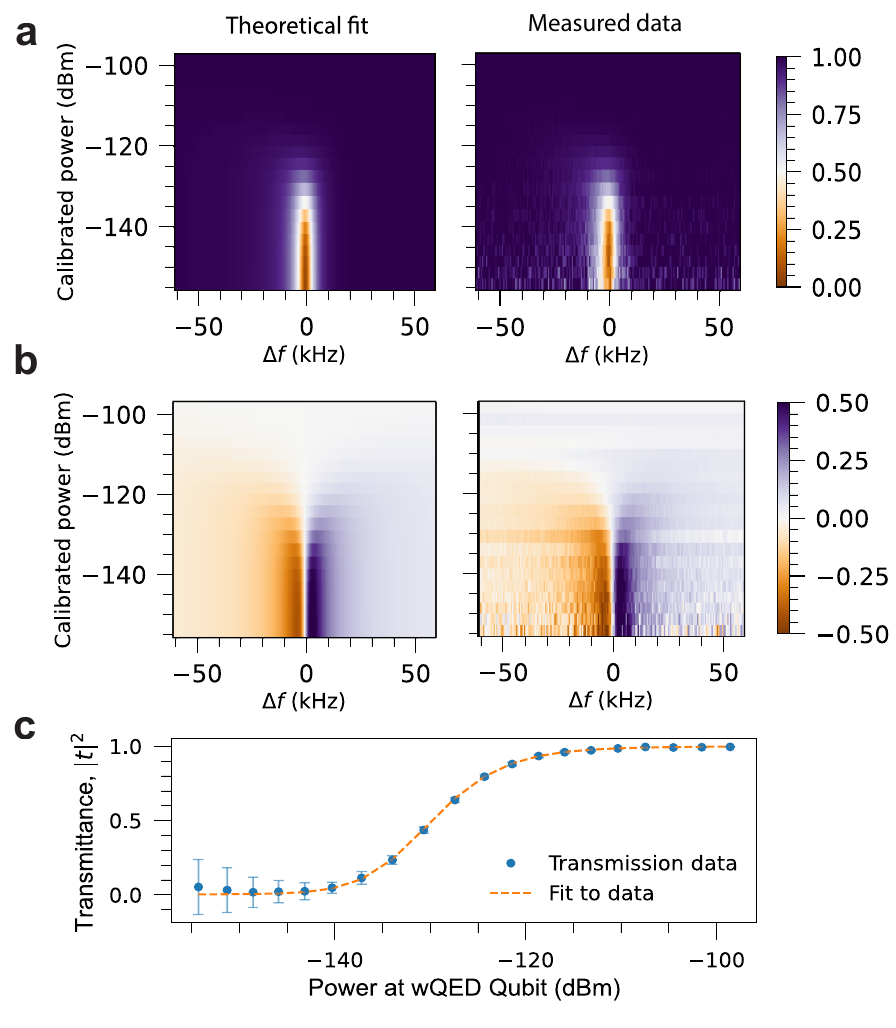}
\caption{(a) Real part of theoretical model and real part of transmission data, in which we observe the qubit perfectly reflecting an input photon at low powers, versus transmitting near unity at higher powers. (b) Imaginary part of theoretical model and imaginary part of transmission data, where we see a phase shift corresponding to the dip in (a). (c) Global fit gives us a transmittance curve at the resonator frequency, which allows us to extract the absolute power at every nominal input power used for the power sweep. }
\label{fig:supp-fig7}
\end{figure}

The wQED device consists of a symmetric qubit coupled to an open transmission line, which functions as a waveguide. The qubit frequency can be tuned with a DC magnetic flux bias to find a flux-insensitive sweet spot, in this case 6.59 GHz. Once tuned to the desired qubit frequency, we perform a 2D scan of the transmission profile by sending in a weak coherent drive at varying input powers. At lower drive powers, the qubit will absorb a single photon from the drive field and emit a photon bidirectionally with a $\pi$ phase shift, thereby perfectly reflecting the single photon through constructive interference in the backwards direction and destructive interference in the forward direction. This results in a dip in the transmission at the resonant frequency. At higher drive powers, the higher number states of the qubit become excited and the resonant transmission increases, since the qubit can only emit up to a single photon. In \cref{fig:supp-fig7} (a) and (b), we note that the real and imaginary parts of the linear power scan exhibit the transmission behavior we expect (the measured data is normalized to transmission with the qubit far detuned). The equation describing transmission for a strongly coupled qubit and waveguide is 
\begin{equation}
t=1-\frac{ \Gamma_1}{2 \Gamma_2} \frac{1-\frac{i \Delta}{\Gamma_2}}{1+\left(\frac{\Delta}{\Gamma_2}\right)^2+\frac{\Omega^2}{\Gamma_1 \Gamma_2}} .
\end{equation}
where $\Gamma_1$ is the decay rate of the qubit into the transmission line, $\Gamma_2$ is the transverse decoherence rate of the qubit, and  $\Omega$ is the drive amplitude in units of Hz \cite{qiu_broadband_2022}. The absolute drive power at the qubit is given by \cite{mirhosseini_cqed_2019}:
\begin{equation}
P=\pi \hbar \omega_i \Omega^2 / 2 \Gamma_1 .
\end{equation}
By performing a global fit of the input power-dependent transmission as shown in \cref{fig:supp-fig7} (c), we can extract the drive amplitude $\Omega$ and the decay rate of the qubit into the transmission line $\Gamma_1$, and thus calculate the absolute power at the input of the wQED device and the Floquet TWPA, which we call reference plane \textbf{A} in Fig 2. a). from the main text. \cite{qiu_broadband_2022}.

\section{System and Intrinsic Amplifier Quantum Efficiency Calculation \label{app:intinsicqe}}

%---------- Fig 8 description ----------%
\begin{figure}
\centering
\includegraphics[width=\columnwidth]{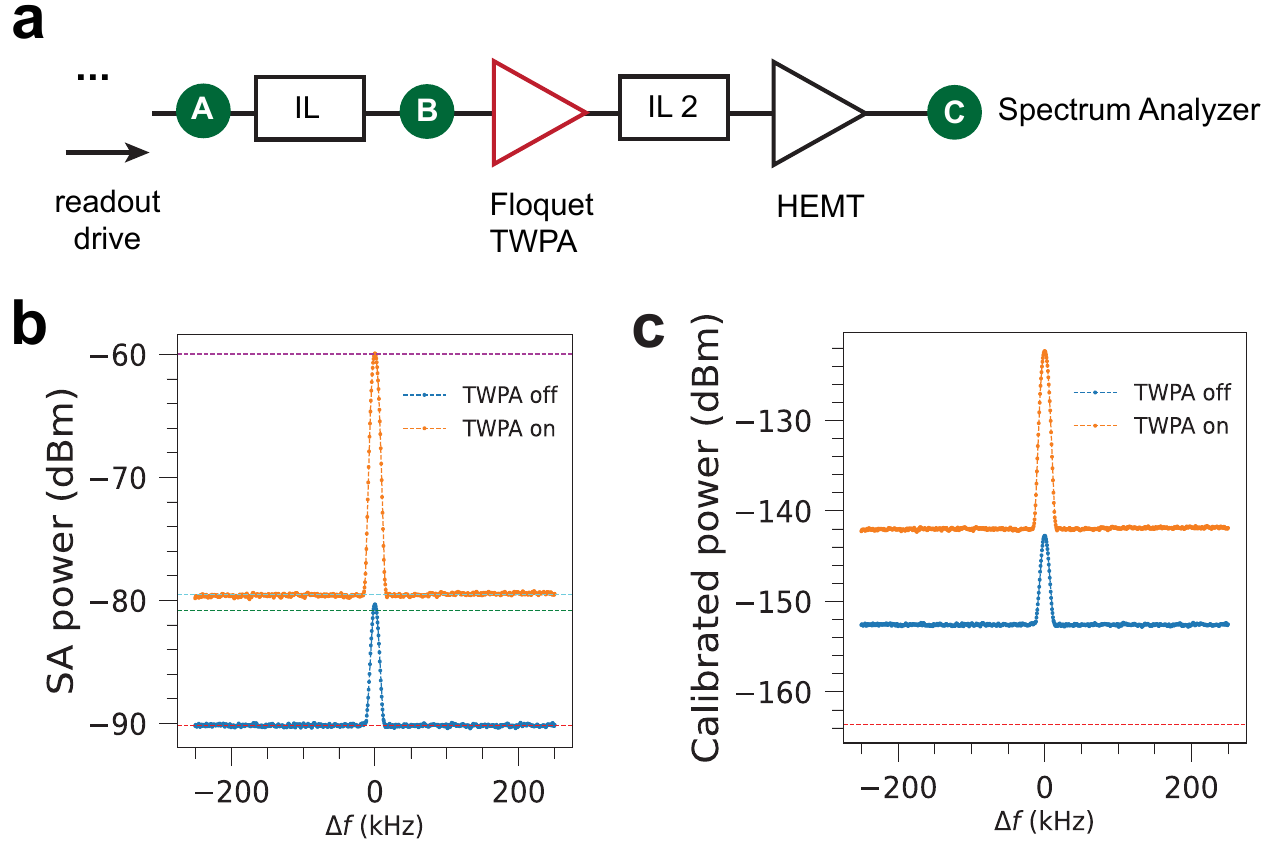}
\caption{(a) Simplified measurement chain with which to derive quantum efficiency if we know the power at reference plane \textbf{A}, obtained from the circuit QED power calibration, or reference plane \textbf{B}, obtained from the waveguide QED power calibration. For the former, we also need to know the insertion loss between \textbf{A} and \textbf{B}, which we estimate from a dunk test. The second insertion loss between the TWPA and the HEMT not necessarily directly measured or calculated, but included in the overall system efficiency. (b) Measured signal with the JTWPA on (orange line) and off (blue line). We fit both curves with a gaussian to extract the respective signal and noise (dark blue, green dashed lines). (c) After obtaining the power calibration, we find a fixed difference factor, defined as the difference between the signal while TWPA pump is off, and the calibrated power while TWPA is off. We can then replot our curves with respect to the standard quantum limit. Here we use the results from the cQED calibration, but the process is exactly the same for the wQED calibration.} 
\label{fig:supp-fig8}
\end{figure}

After obtaining absolute power reference at the input of the Floquet TWPA, we perform noise measurements when the Floquet TWPA is either turned on and off to evaluate system and intrinsic quantum efficiency. Consider the simplified readout chain diagram shown in \cref{fig:supp-fig8}(a), in which we combine and abstract various circuit components to several functional blocks, without loss of generality: the JTWPA and HEMT amplifiers, each with added noise and gain, as well as insertion loss before and between the amplifiers. We define reference planes \textbf{A}-\textbf{C} for clarity. From the cQED calibration, we can obtain the absolute power value at plane \textbf{A}, while from the wQED calibration, we can obtain the absolute power value at plane \textbf{B}. The insertion loss between these points can be estimated from a dunk test of the components (cables, circulator, and directional coupler) in liquid nitrogen. We will use the following notations: $S_X$ and $N_X$ refer to the calibrated power of the signal and noise floor at the reference plane \textbf{X}, and additional subscripts $_{\mathrm{amp,i}}$, and $_{\mathrm{sys,on(off)}}$ denote values concerning the intrinsic performance of the TWPA itself and full system performance while the TWPA is on (off), respectively. For instance, $S_{A,\mathrm{sys,off}}$ corresponds to the signal power at reference point \textbf{A} while the JTWPA is turned off.

From the room temperature measurements at \textbf{C} and power calibration at \textbf{A}, we extract the system gain, downstream gain (when TWPA is turned off), and amplifier gain in linear scale:
\begin{equation}
\begin{aligned}
G_{\mathrm{sys, on} } &= S_{D, \mathrm{sys,on}} / S_{A, \mathrm{sys,on}} \\
G_{\mathrm{sys,off}} &= S_{D, \mathrm{sys,off}} / S_{A, \mathrm{sys,off}} \\
G_{\mathrm{amp,i}} &= S_{D, \mathrm{sys,on}} / S_{D, \mathrm{sys,off}}. 
\end{aligned}
\end{equation}

From the gain values, we can then extract the noise temperature from noise power by dividing by the Boltzmann constant $k_B$ and the resolution bandwidth $B=10\,$kHz on the spectrum analyzer:
\begin{equation}
T_{\mathrm{sys,on(off)}} = \frac{N_{A,{\mathrm{sys,on(off)}}}}{k_B B} =  \frac{N_{D,\mathrm{sys, on(off)}} / G_{\mathrm{sys, on(off)}}}{k_B B}.
\end{equation}

The system measurement efficiencies with the TWPA on and off can thus be calculated from the noise temperatures using \cref{eq:qesysdef}, repeated here for convenience.
\begin{equation}
\eta_{\mathrm{sys, on(off)}} = \frac{\hbar \omega}{k_B T_{\mathrm{sys, on(off)}} }
\end{equation}

Finally, we calculate the amplifier intrinsic quantum efficiency from the extracted system noise metrics. The total noise photon numbers when the TWPA is turned on and off (including the $1/2$ vacuum noise photon) referred at the amplifier input are

\begin{equation}
    n_{\mathrm{sys,on(off)}} =  \frac{k_B T_{\mathrm{sys, on(off)}} }{\hbar \omega} = 1/\eta_{\mathrm{sys, on(off)}},
\end{equation}

and the added noise photon number from the Floquet TWPA is thus
\begin{equation}
A_{\mathrm{amp,i}} = (n_{\mathrm{sys,on}}- 1/2) - \left(\frac{n_{\mathrm{sys,off}}- 1/2}{G_{\mathrm{amp,i}}} \right),
\end{equation}
resulting in the normalized intrinsic efficiency \cite{peng_floquet_2022}
\begin{equation}
\begin{aligned}
\Tilde{\eta}_{\mathrm{amp,i}} &= {(1 + 2A_{\mathrm{amp,i}})}^{-1}  / \eta_{\mathrm{ideal}}(G_{\mathrm{amp,i}}) \\
&= \frac{2 - 1/G_{\mathrm{amp,i}} }{1+2A_\mathrm{amp,i}} \\
&= {\left(\frac{2}{\eta_{\mathrm{sys,on}}} - \frac{2}{G\eta_{\mathrm{sys,off}}} + \frac{1}{G}\right)}^{-1}\!\!\Big/ \, \eta_{\mathrm{ideal}}(G),
\end{aligned}
\end{equation}
same as \cref{eq:qeintrinsicdef} in the main text. 

\section{Circuit Quantum Electrodynamics (cQED) Power Calibration \label{app:qubitcal}}

In \cref{sec:sysmeaseff} and \cref{fig3}, we performed circuit QED power calibration QED method \cite{gambetta_mid_2006,macklin_a_2015} to extract the system measurement efficiency in the qubt readout measurement setup. This method utilizes the effect that the AC stark shift and the measurement-induced dephasing rate of a qubit is dependent on the photon number population of the resonator it is dispersively coupled to. Thus, by simultaneously fitting the dependence of the frequency shift and the measurement-induced dephasing to the varying drive power, or to the unique function dependence on the drive frequency detuning of a weak measurement tone around the resonator, one can reliably calibrate the photon population in the resonator and consequently the absolute power reference for the measurement tone at the input of the qubit chip at the mixing chamber.

%---------- Fig 9 description ----------%
\begin{figure}[bhtp]
\centerline{\includegraphics[width=\linewidth]{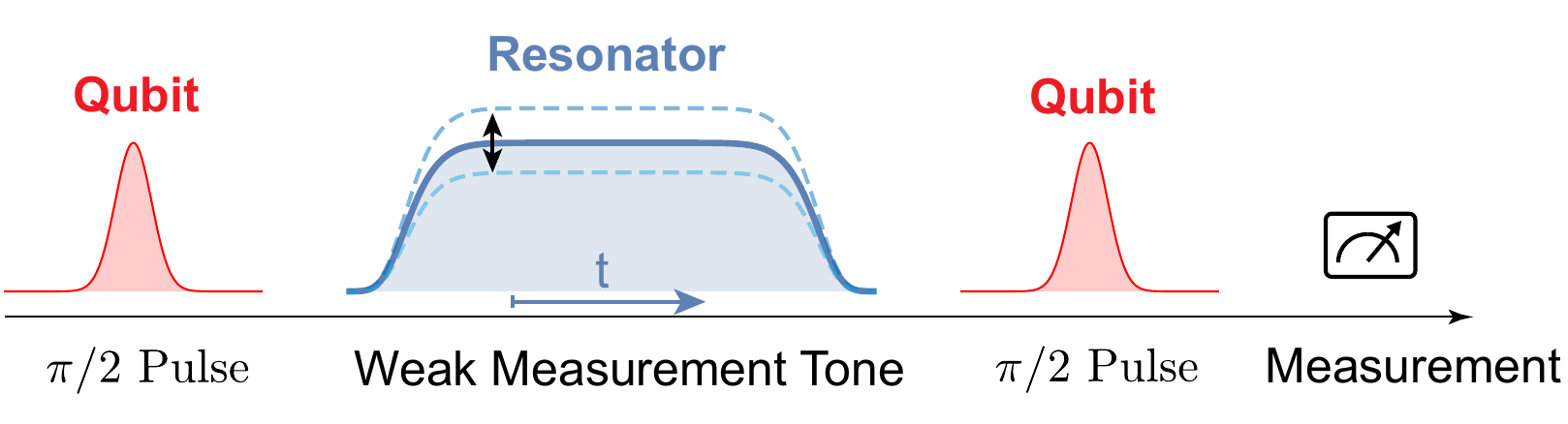}}
\caption{Pulse Sequence of the circuit-QED power calibration method. In each Ramsey-like measurement, the duration of the flat-top Gaussian weak measurement tone in between the two $\pi/2$ qubit pulses is varied to extract the measurement-induced dephasing rate and AC Stark shift from the decaying sinusoidals at a particular drive tone power. Such measurements are then repeated at varying input drive tone power for power calibration fitting.}
\label{fig:supp-fig9}
\end{figure}

\Cref{fig:supp-fig9} shows the pulse sequence of the aforementioned circuit QED power calibration method. This sequence is similar to the well-known Ramsey measurement, except that an additional weak measurement pulse with varying power is introduced in between the two $\pi/2$ pulses instead of simply idling as in the Ramsey sequence. With a drive Hamiltonian $\hat{\mathcal{H}}_{d} = \epsilon_d\hat{a}e^{i\omega_d t} + H.c.$, in which $H.c.$ stands for the Hermitian conjugate of the preceding term and $\omega_d$ is the frequency of the drive tone, the steady-state AC stark shift $\omega_{AC}$ and the measurement-induced dephasing rate $\Gamma_m$ can be expressed as \cite{gambetta_mid_2006}
\begin{equation}
    \begin{aligned}
        \omega_{AC} &= 2\chi\Re{\alpha_e^*\alpha_g},  \\
        \Gamma_m &= 2\chi\Im{\alpha_e^*\alpha_g},  \quad \
    \end{aligned} \label{eq:powercalformula}
\end{equation}
in which
\begin{equation}
        \alpha_{e(g)} = \frac{-j\epsilon_d}{\kappa/2 + j(\Delta_r \pm \chi)}.
\end{equation}

In the above formulae, $\Delta_r = \omega_{r0} - \omega_d$ is the resonator drive detuning, $\chi$ is the dispersive shift between the qubit and the resonator, $\omega_{r0}$ is the bare (uncoupled) resonant frequency of the resonator, and $\kappa = \kappa_{\mathrm{int}} + \kappa_{\mathrm{ext}}$ is the total or the loaded decay rate of the resonator. Utilizing the input-output theory, we can express the input power in the feedline with the external coupling rate $\kappa_{\mathrm{ext}}$ as
\begin{equation}
    P_{\mathrm{in}} = \hbar\omega_d\epsilon_d^2/\kappa_{\mathrm{ext}}. \label{eq:powerformula}
\end{equation}

%---------- Fig 10 description ----------%
\begin{figure*}[ht]
\centering
{\includegraphics[width=\linewidth]{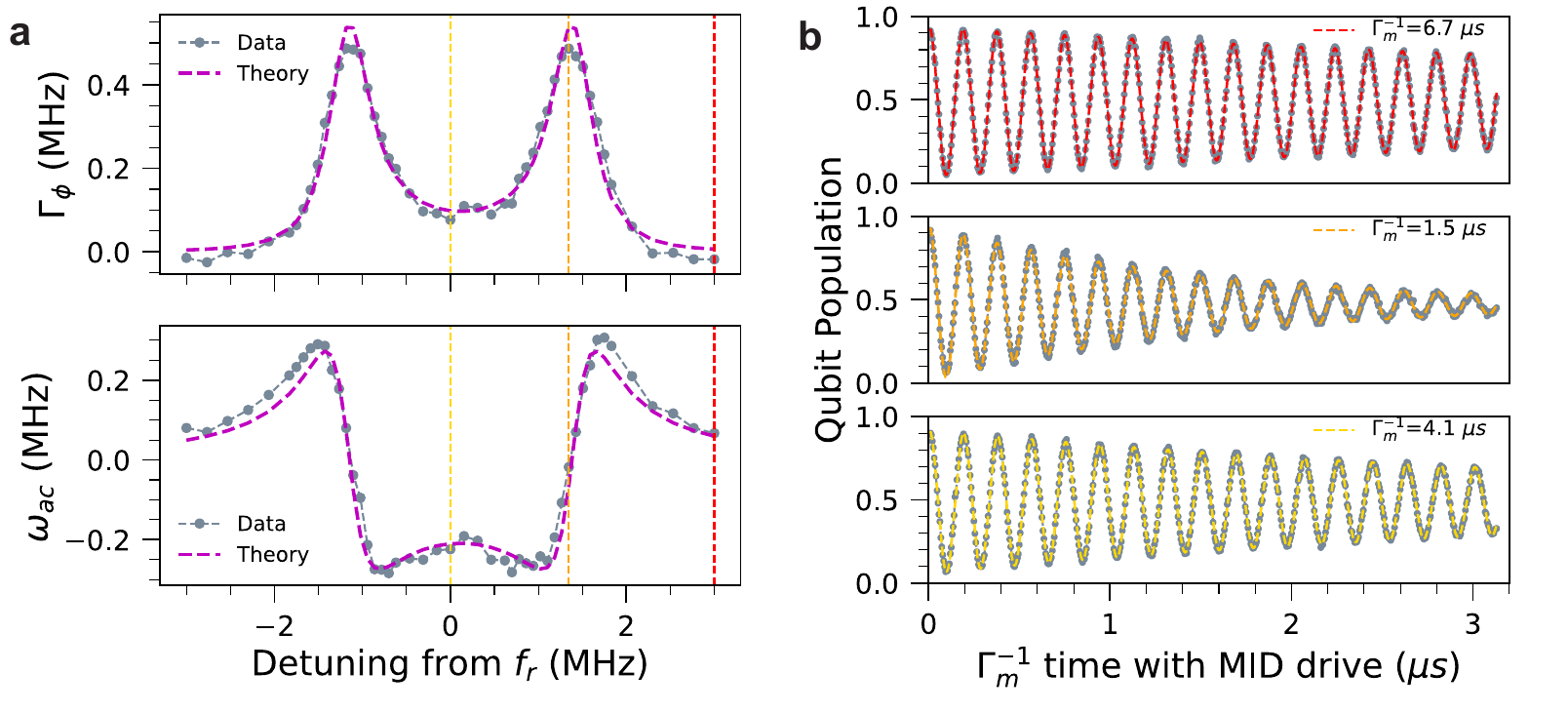}}
\caption{(a) The measurement-induced dephasing rate $\Gamma_m$ and AC Stark Shift $\omega_{AC}$ as a function of the detuning $\Delta_r=\omega_{r0}-\omega_d$. (b) Measurement-induced dephasing Measurements (\cref{fig:supp-fig9}) as a function of the dephasing pulse duration at far-detuned (top), near $\ket{1}$ state resonance (middle), and at bare resonance (bottom), respectively.}
\label{fig:supp-fig10}
\end{figure*}

\Cref{fig:supp-fig10}(a) shows the measured AC Stark Shift $\omega_{AC}$ and measurement-induced dephasing rate $\Gamma_m$ as a function of the drive detuning $\Delta_r$. \Cref{fig:supp-fig10}(b) shows three representative linecuts, in which the weak measurement tone is far-detuned, at $\ket{1}$ state resonance, and bare-resonance $\omega_{r0}$, respectively. In the far-detuned case (\cref{fig:supp-fig10}(b) top), the resonator is barely populated by the drive and thus has minimal excess dephasing. In contrast, the measurement-induced dephasing is maximized around the $\ket{1}$ resonance (\cref{fig:supp-fig10}(b) bottom). 

%---------- Fig 11 description ----------%
\begin{figure}[bhtp]
\centerline{\includegraphics[width=\linewidth]{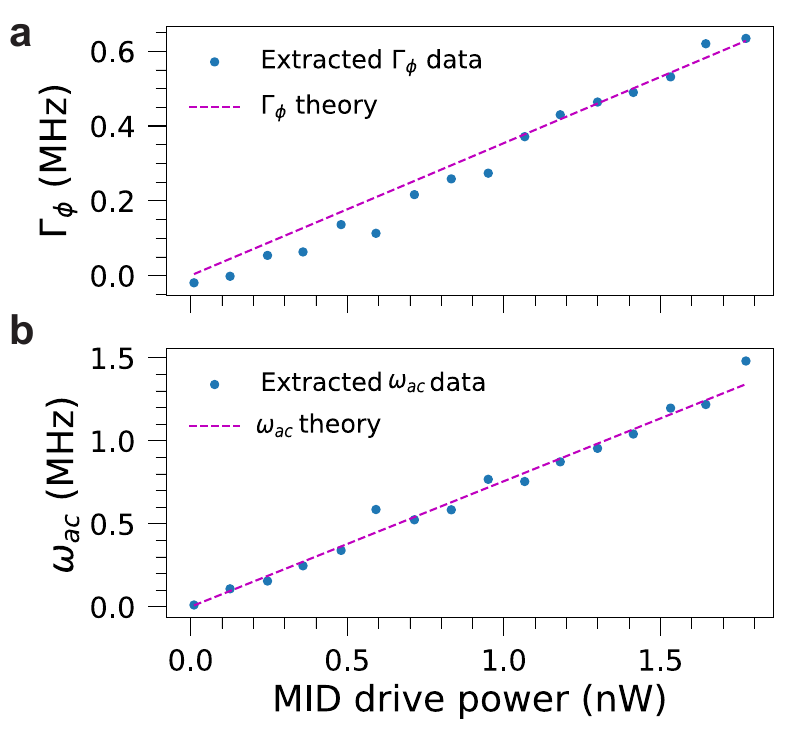}}
\caption{(a) The measurement-induced dephasing rate $\Gamma_m$ (a) and AC Stark Shift $\omega_{AC}$ (b) as a function of the input DAC power at room temperature.}
\label{fig:supp-fig11}
\end{figure}

\Cref{fig:supp-fig11} shows the measured AC Stark Shift $\omega_{AC}$ and measurement-induced dephasing rate $\Gamma_m$ as a function of the DAC input power. We note the bottom x-axis values $P_{\mathrm{DAC}}$ (or MID pulse drive power) here are referenced at the input going into the fridge at room temperature and therefore do not correspond to the power at the chip to be calibrated, but instead act as a scaling reference for power linearity and fitting. The top x-axis corresponds to the power-calibrated number of photons in the resonator after fitting. The dashed-line fits in \cref{fig:supp-fig11} is obtained by applying the experimentally extracted parameters $\chi$, $\omega_{r0}$, and $\kappa$ presented in \cref{tab:powercalqubitparams} and using $c_\epsilon = \epsilon_d / \sqrt{P_{\mathrm{DAC}}}$ as the only free fitting parameter. We can then extract the photon number population and thus input measurement tone power at each nominal DAC input power reference from the simultaneous fit for $\omega_{AC}$ and $\Gamma_m$ for use in noise characterization in the main text.

\begin{table}[t]
\begin{center}
\caption{Circuit QED parameters of the qubit-resonator system used for power calibration.}
\vspace{10pt}
\begin{tabular}{cccccccc}
\toprule
 $\omega_q$ & $\omega_{r0}$ & $\chi$ & $\kappa$ &  $\kappa_{\mathrm{ext}}$\\
(GHz) &(GHz) & (MHz) & (MHz)  & (MHz) \\ \midrule
  $5.769$ & $6.505$ & $1.296\pm0.03$ & $0.572\pm 0.04$ & $0.458\pm0.002$ \\  %\hline
\bottomrule
\end{tabular}
\label{tab:powercalqubitparams}
\end{center}
\end{table}

%---------- Fig 12 description ----------%
\begin{figure}[bhtp]
\centerline{\includegraphics[width=0.8\linewidth]{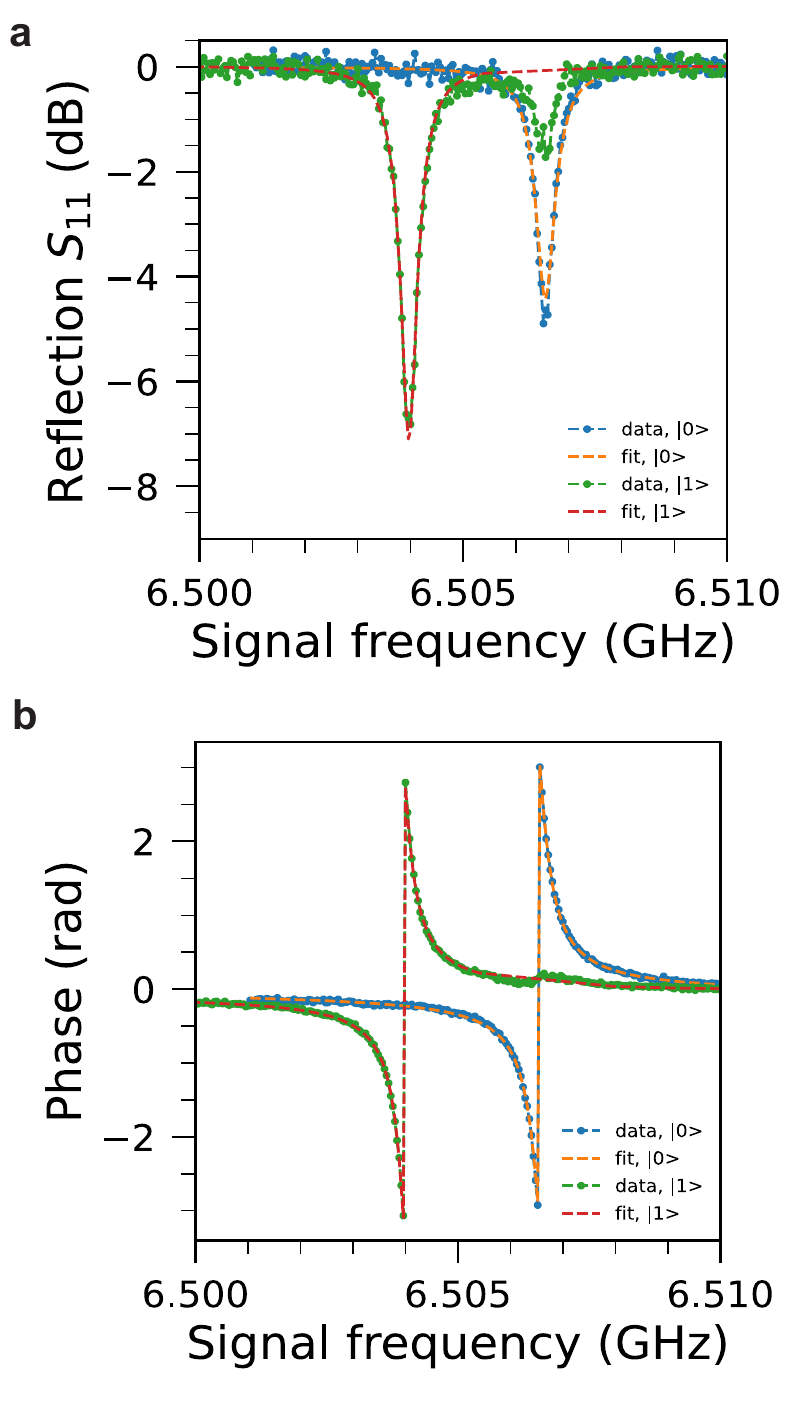}}
\caption{Resonator spectroscopy performed with qubit in ground (blue data, orange fit) versus excited (green data, red fit) states. Panel a) is the reflection and panel b) is the phase.}
\label{fig:supp-fig12}
\end{figure}

For the fits described above, we obtain the $\chi$ and $\kappa$ values from resonator spectroscopy. Since this is a reflection measurement, it is possible to determine the external and internal losses by measuring S parameters and phase of the resonator features, shown in Fig.~\ref{fig:supp-fig12}. We utilize the following model for asymmetric resonators as follows \cite{probst_res_2015, khalil_res_2012}:

\begin{equation}
S_{11}(f)=a e^{i \alpha} e^{-2 \pi i f \tau}
\left[\frac{\left(2 Q_l /\left|Q_c\right|\right) + 2 i Q_l\left(f / f_r-1\right)}{1 - 2 i Q_l\left(f / f_r-1\right)}\right].
\end{equation}

From fitting the resonator dips while the qubit is in the ground and excited states, we extract the values shown in Table.~\ref{tab:powercalqubitparams}. We then input the $\chi$ and $\kappa$ values into the theoretical fit for our measurement-induced dephasing frequency and power sweeps in \cref{fig:supp-fig10} and \cref{fig:supp-fig11}. This allows us to extract a power factor $\epsilon_d$ from both sweeps at a chosen MID drive power, obtaining a difference of $5.293\%$ between the two obtained factors; this leads to a difference of 0.45 dB in the absolute power calibration, -142.49 dBm and -142.04 dBm respectively. For the cQED power calibration in this work, we choose the value obtained from the frequency detuning MID sweep \cref{fig:supp-fig10} since it is a direct, well-averaged measurement over 49 different frequency points at our desired drive power, as opposed to a single frequency for the power-dependent MID sweep in \cref{fig:supp-fig11}.

\section{Full Measurement Setup}%

%---------- Fig 13 description ----------%
\begin{figure*}[ht]
\centering
\includegraphics[width=\linewidth]{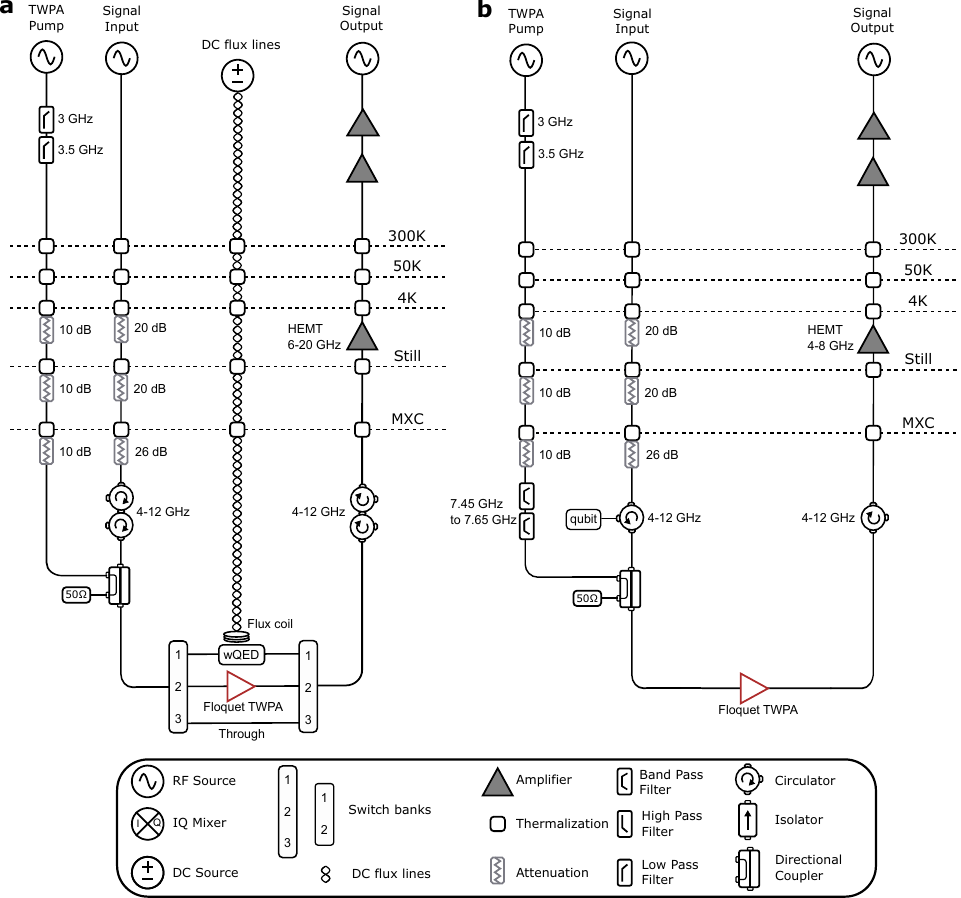}
\caption{ (a) Fridge diagram for the setup to characterize gain, insertion loss, and intrinsic quantum efficiency performance metrics of the Floquet TWPA, including a through line and wQED device in parallel with the Floquet TWPA. (b) Fridge diagram for the setup to characterize Floquet TWPA performance in a qubit readout chain. The transmon qubit is measured in reflection and used to perform a power calibration (described in Appendix D) to obtain the measurement efficiency of the full system.}
\label{fig:supp-fig13}
\end{figure*}

We characterize the Floquet TWPA in this paper with two setups: first, in a switchbank optimized for extracting TWPA metrics referred to a through line, and second, in a qubit readout chain optimized for high system efficiency. The two setups are depicted in \cref{fig:supp-fig13}. For the first setup in \cref{fig:supp-fig13} (a), the gain and insertion loss measurements, as well as the transmission measurements of the wQED device, are performed with a vector network analyzer (VNA; Hewlett Packard 8720ES). For noise rise experiments, we utilize an RF generator source (Windfreak; SynthHD PRO v2) for a clean signal tone with minimal sidebands, and inspect the SNR improvement on a spectrum analyzer (SA; Agilent E4440A). The TWPA pump tone is provided by a separate RF generator (Windfreak; SynthHD PRO v2) at room temperature, and couples into the signal line with a directional coupler (Marki; C20-0226) at cryogenic temperatures. The Floquet TWPA is situated in parallel with the wQED device and through line in a Radiall 6-port switchbank (Radiall R591762600); we utilize non-magnetic cables of the same length and make within the switchbank in order to provide the best possible reference for the line with the Floquet TWPA. The cables in the through line are connected with a SMPM barrel, as a close analog to the adapters used to interface between the SMA-connectorized non-magnetic cables and the TWPA package input with SMPM connectors. Before the directional coupler and after the switchbank, we have 4-12 GHz isolators and circulators (LNF; ISISC4-12A, CICIC4-12A) to provide directionality and minimize the effect of downstream reflections. The signal goes through a HEMT amplifier (LNF; LNC-6-20D), is further amplified by two room temperature amplifiers (Mini-Circuits ZX60-05113-LN+), and reaches room temperature readout, which in turn can go to port 2 of the VNA for transmission measurements, or the SA for noise rise experiments. 

The setup in \cref{fig:supp-fig13} (b) consists of a simple qubit readout chain, with a transmon qubit measured in reflection, utilizing the same fridge attenuation and TWPA pump source as in (a). To minimize heating from the TWPA pump tone, we add two bandpass filters (MCL VBF-7500+ 7450 to 7650 MHz BPF) to the pump line at the mixing chamber stage. Since this measurement chain is optimized for high system efficiency, we minimize the insertion loss between the qubit and the Floquet TWPA by using a single junction circulator (LNF; CICIC4-12A) along with a low-loss aluminium cable. In addition, we reduce the output circulator to a single junction circulator, and use a HEMT amplifier (LNF; LNC4-8C) that has a low noise temperature of 1.5 K. For transmission experiments, we likewise obtain data with the VNA. For qubit readout, power calibration, and noise rise experiments, the input signal tone is provided by a digital-to-analog-converter (DAC) pin of an FPGA (AMD Zynq UltraScale+ RFSoC ZCU111, Xilinx) running the Quantum Instrumentation Control Kit (QICK) software \cite{qick}. The output tone is read out either by the native analog-to-digital converter (ADC) of the FPGA for qubit experiments, or with the SA for noise rise experiments. 

\bibliography{main}

\end{document}